\documentclass[aps,twocolumn,showpacs]{revtex4}
\usepackage{epsfig}
\usepackage{graphicx}
\usepackage{amsmath}
\usepackage{mathrsfs}
\usepackage{dcolumn}

\newcommand{\beq}{\begin{eqnarray}}
\newcommand{\eeq}{\end{eqnarray}}
\newcommand{\be}{\begin{equation}}
\newcommand{\ee}{\end{equation}}
\newcommand{\bey}{\begin{eqnarray}}
\newcommand{\eey}{\end{eqnarray}}
\newcommand{\ba}{\begin{array}}
\newcommand{\ea}{\end{array}}
\newcommand{\bi}{\begin{itemize}}
\newcommand{\ei}{\end{itemize}}
\newcommand{\bem}{\begin{enumerate}}
\newcommand{\eem}{\end{enumerate}}
\newcommand{\bw}{\begin{widetext}}
\newcommand{\ew}{\end{widetext}}
\newcommand{\ra}{\rangle}
\newcommand{\la}{\langle}

\newcommand{\E}{{\cal E}}

\newcommand{\mP}{{\mathscr{P} }}
\newcommand{\oa}{\overline{\alpha }}

\begin{document}

 \title{
 Entanglement-induced Decoherence and Energy Eigenstates}
 \author{Wen-ge Wang$^{1,2,3}$, Jiangbin Gong$^{1,6}$, G.~Casati$^{1,4,5}$, and Baowen Li$^{1,6}$}
 \affiliation{
 $^1$Department of Physics and Centre for Computational Science and Engineering,
 National University of Singapore, 117542, Republic of Singapore
 \\ $^{2}$Department of Modern Physics, University of Science and Technology of China,
 Hefei 230026, China
  \\ $^{3}$Department of Physics, Southeast University, Nanjing 210096, China
 \\ $^{4}$Center for Nonlinear and Complex Systems, Universit\`{a}
  degli Studi dell'Insubria, Via Valleggio 11, 22100 Como, Italy
 \\ $^5$CNR-INFM and Istituto Nazionale di Fisica Nucleare, Sezione di Milano, Italy
 \\  $^6$NUS Graduate School for Integrative Sciences and Engineering, Singapore
 117597, Republic of Singapore
 }

 \date{\today}

 \begin{abstract}
Using recent results in the field of quantum chaos we derive
explicit expressions for the time scale of decoherence induced by
the system-environment entanglement. For a generic
system-environment interaction and for a generic quantum chaotic
system as environment, conditions are derived for energy eigenstates
to be preferred states in the weak coupling regime. A simple model
is introduced to numerically confirm our predictions. The results
presented here may also help understanding the dynamics of quantum
entanglement generation in chaotic quantum systems.
 \end{abstract}

 \maketitle

 \section{Introduction}

 Real physical systems are never isolated from the surrounding
 world and, as a consequence, nonclassical correlations (entanglement)
 are established between the system and the
environment. This process, which leads to decoherence, has a
fundamental interest since it contributes to the understanding of the
emergence of classicality in a world governed by the laws of quantum
mechanics \cite{Zurek81}.

 Remarkably, different states may decohere at drastically different rates,
 and a small fraction of them may be particularly stable
 under entangling interaction with the environment \cite{Zurek81,BHS01}.
 Such states are ``preferred states" of the system under the influence
 of the environment. (They are also called ``pointer states", a name given
 for the states of pointers of measurement apparatus in the study of the
 measurement problem \cite{Zurek81}.)  This concept is important
 in understanding what states will naturally emerge from a quantum
 system subject to decoherence.

As discussed by Paz and Zurek in an interesting paper \cite{PZ99},
depending on the type and strength  of the system-environment
interaction, different preferred states may arise during the
decoherence process.
 In particular they consider the case of an adiabatic environment modeled by a quantum scalar
field and interacting weakly with a system
 through a given type of coupling. In such situation they show
 that energy eigenstates
 are good preferred states  and hence are
 the natural representation of the quantum system.
 Notice that the adiabatic environment does not change the population of energy eigenstates
 of the system, implying infinite relaxation time.

 In realistic situations the environment is often nonadiabatic, with finite
 relaxation time.
 In this case, the relation between the relaxation time and the decoherence time is crucial
 and, typically, only when the former is much longer than the latter,
 energy eigenstates can be good preferred states.
 While a rough estimate of the relaxation time can be obtained via Fermi's golden rule,
 for the decoherence time the situation is more complex. Indeed
 in case of generic type of coupling and generic environment, it is hardly possible to
 obtain estimates by employing
 the master-equation approach used in \cite{PZ99}.

 In this paper we propose an alternative approach which takes advantage of recent progress
 in the field of quantum chaos: namely random matrix theory and the so-called fidelity
 \cite{wang,fid-chao,CT02,GPSZ06,CDPZ03} which
 is a measure of the stability of the quantum motion under system perturbations.
 Using this approach we can estimate - via the fidelity decay of the
 environment - the decoherence time of the system
 for {weak}, generic system-environment couplings and for a broad class of environments.
 In particular, we need not restrict ourselves to the Markovian regime
 (in contrast to master-equation approach) where the bath correlation decay is faster than
 decoherence.
 Moreover, as an important result of ours,
 by modeling the environment by a chaotic quantum system,
 we determine a critical border in the coupling strength below which
 energy eigenstates are shown to be preferred states, while above this border, the relaxation
 time and decoherence time have the same scaling with the coupling strength and
 therefore no definite statements can be made. We also
 present numerical results which confirm the above picture.

 \section{General theory}

 Let us consider
 a quantum system $S$ with a discrete spectrum,
 weakly coupled to a second quantum system $\E$ as its environment .
 The total Hamiltonian is:
 \be H = H_S + \epsilon H_I + H_{\E}, \ee
 where $H_S$ and $H_{\E}$ are the Hamiltonians of $S$ and $\E$ respectively, and
 $\epsilon H_I$ is the weak interaction Hamiltonian ($\epsilon \ll 1$).
 The time evolution of the whole system is given by $|\Psi_{S\E}(t)\ra
 = e^{-iHt/ \hbar } |\Psi_{S\E}(0)\ra $.
 The initial state is set as a product state
 $|\Psi_{S\E}(0)\ra =|\psi_S(0) \ra |\phi_{\E}(0)\ra $.
 The reduced density matrix $\langle\alpha|\rho^{re}(t)|\beta\rangle =
 \langle\alpha|{\rm Tr}_{\E} \rho (t)|\beta\rangle$ is obtained by
 tracing over the environment. 

Consider first the case  with initial state $|\psi_S(0)\ra  = |\alpha \ra $, where
$|\alpha \ra $ denotes an energy eigenstate of $H_S$ with
eigenenergy $E_{\alpha }$. We define the projection operators
$|\alpha \ra \la \alpha | \otimes 1_{\E}$ and $ \mP_{\oa} \equiv
\sum_{\beta \ne \alpha } |\beta \ra \la \beta | \otimes 1_{\E}$
where $1_{\E}$ is the identity operator for the environment degrees
of freedom. The whole Hilbert space can then  be decomposed into two
orthogonal subspaces, leading to:
 \be |\Psi_{S\E}(t)\ra  = e^{-iE_{\alpha }t/ \hbar } |\alpha \ra |\phi^{\E}_{\alpha }(t)\ra
 + \epsilon |\chi_{\oa }(t) \ra , \label{Psi-t} \ee
 where $ \epsilon |\chi_{\oa }(t) \ra \equiv \mP_{\oa} |\Psi_{S\E}(t)\ra $.
  The small parameter $\epsilon$ is introduced to account for the fact that,
  in case of weak coupling, the second term
  in Eq.~(\ref{Psi-t}) remains small inside some initial time interval (see below).
 The normalization of $|\Psi_{S\E}(t)\ra$ in Eq.~(\ref{Psi-t})
 is unity to the first order in $\epsilon$.

 A simple derivation shows that the evolution of the two terms in Eq. (\ref{Psi-t}) is given
 by the coupled equations,
 \bey \label{phi-t} i\hbar \frac{d}{dt} |\phi^{\E}_{\alpha }(t)\ra
 =  H^{eff}_{\E \alpha }  |\phi^{\E}_{\alpha }(t)\ra
  + \epsilon^2 e^{iE_{\alpha }t/ \hbar } \la \alpha |H_I|\chi_{\oa }(t) \ra ,
 \\  \label{eta-t}  i\hbar \frac{d}{dt} |\eta_{\oa }(t) \ra =
 \exp \left [ -\frac i{ \hbar}(E_{\alpha } - H_{\oa} ) t
 \right ] \mP_{\oa} H_I |\alpha \ra |\phi^{\E}_{\alpha }(t)\ra , \eey
 where $H^{eff}_{\E \alpha } \equiv \epsilon H_{I\alpha } + H_{\E} $,
 $|\eta_{\oa }(t) \ra  \equiv  \exp \left ( i H_{\oa}  t/ \hbar \right ) |\chi_{\oa }(t) \ra $,
 $H_{I\alpha } \equiv \la \alpha | H_I | \alpha \ra $, and
 $H_{\oa} \equiv \mP_{\oa} H \mP_{\oa} $.

 It is evident from Eq.~(\ref{Psi-t}) that $\epsilon^2 \la \chi_{\oa}|\chi_{\oa}\ra $
 gives the population that has leaked to the subspace associated with $\mP_{\oa}$.
 In the case of weak coupling, this population leakage is initially very small.
 More precisely we have that $ \epsilon^2 \la \chi_{\oa}|\chi_{\oa}\ra \ll 1$
 up to
 times $ t\ll \tau_E$, where $\tau_E$ is the relaxation time of the system.
Then the second term in Eq.~(\ref{phi-t}) can be safely neglected
and, as a result, the environment is in the state
$|\phi^{\E}_{\alpha }(t)\ra
 \approx e^{-itH^{eff}_{\E \alpha } / \hbar} |\phi_{\E}(0)\ra $
while the system remains in the eigenstate $|\alpha\rangle$ with a definite phase evolution.

Consider now as the initial state a superposition of energy eigenstates
 $ |\psi_S(0)\ra = \sum_{\alpha } C_{\alpha } |\alpha \ra $.
 {As in Eq.~(\ref{Psi-t}) we can write:}
 \be |\Psi_{S\E}(t)\ra =
 \sum_{\alpha } e^{-iE_{\alpha }t/ \hbar } C_{\alpha } |\alpha \ra |\phi^{\E}_{\alpha }(t)\ra +
 \epsilon |\chi (t)\ra , \label{Psi-t2} \ee
 where the term $|\chi (t)\ra =  \sum_{\alpha } C_{\alpha } |\chi_{\oa}(t) \ra $
 contains now contributions of transitions between different energy eigenstates.
 Note that the first term on the right hand side of Eq.~(\ref{Psi-t2}) may be highly
 entangled even when $\epsilon |\chi (t)\ra$ is small.
 {For $t\ll \tau_E$, the term $\epsilon |\chi (t)\ra$ in Eq.~(\ref{Psi-t2})
 can be neglected and, by tracing out the environment, the reduced density matrix of the system writes:}
 \bey  \rho^{re}_{\alpha \beta } =\langle\alpha|{\rm Tr}_{\E} \rho (t)|\beta\rangle
  \simeq  e^{-i(E_{\alpha }-E_{\beta})t/ \hbar }
 C_{\alpha } C_{\beta }^* f_{\beta \alpha }(t), \label{off-rho} \hspace{0cm} \eey
 where $f_{\beta \alpha }(t)  \equiv  \la \phi^{\E}_{\beta }(t)|\phi^{\E}_{\alpha }(t)\ra$
 satisfies
 \bey f_{\beta \alpha }(t)  \approx  \la \phi_{\E}(0)|e^{it \left ( H^{eff}_{\E \alpha } + \epsilon V
 \right )/\hbar } e^{-itH^{eff}_{\E \alpha } / \hbar} |\phi_{\E}(0)\ra , \
 \label{ft}
 \\ {\rm with} \ \
     V  \equiv   H_{I\beta } - H_{I\alpha } =
  \la \beta | H_I | \beta \ra - \la \alpha | H_I | \alpha \ra .
 \label{V-define}  \eey
Significantly, the quantity $f_{\beta \alpha }(t)$ is simply the
``fidelity amplitude" of the environment
 associated with the two slightly different
 Hamiltonians $H^{eff}_{\E \alpha }$ and $ (H^{eff}_{\E \alpha } + \epsilon V)$.
 For non-constant $V$, this
 fidelity amplitude usually decays with time for $\alpha\ne \beta$, then $\rho^{re}_{\alpha\beta}$
 also decays and therefore decoherence sets in.
 In case of constant $V$ in Eq.~(\ref{V-define}), Eq.~(\ref{ft}) gives
 $|f_{\beta \alpha}(t)|\approx 1$, hence, there is no decoherence induced by the environment
 in the first order perturbation theory discussed above.
Notice that Eqs.~(6) and (7) become exact in the particular case in
which the coupling $H_{I}$ commutes with $H_{S}$. This case has been
considered in \cite{GPSS04}.

 \begin{figure}[!t]
  \includegraphics[width=\columnwidth]{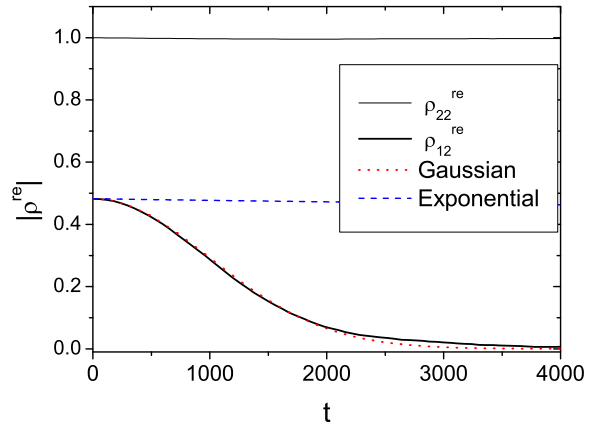}
   \vspace{-0.4cm}
    \caption{
Time dependence of the  elements of the reduced density matrix of
the qubit system $S$. Here $N=200$ and the coupling strength
$\epsilon =5\times 10^{-4}$ is much smaller than the perturbative
border $\epsilon_p \sim 0.04$ numerically computed from
Eq.~(\ref{per-bor}).
The upper thin solid  curve gives the diagonal matrix element
$\rho_{22}^{re}(t)$ for an initial state with $|\psi_S(0)\ra =|2 \ra
$. The lower thick solid curve gives the off diagonal matrix element
$\rho_{12}^{re}(t)$ for an initial state which is a  product state
in which the system S is in an energy superposition state while the
environment is in a randomly chosen state. The dotted and dashed
curves give the theoretical Gaussian decay Eq.~(\ref{Gaussian})
and the exponential decay Eq.~(\ref{ft-fgr}), respectively.
            } \label{fig-1}
         \end{figure}

 It turns out, from the above considerations, that for weak, but generic coupling,
 energy eigenstates of $H_{S}$ play a special role in the sense that, only for
 energy superposition states (not single eigenstates), the decoherence
 process is associated with the instability of the environment under perturbation (fidelity decay).

 \section{A model with the environment modelled by a quantum chaotic system}

Let us now turn to an explicit estimate of the decoherence time of the system.
 To this end we can directly apply recent results on fidelity decay
 \cite{fid-chao,CT02,wang,GPSZ06} which, as  shown below, allows us to
 estimate the decoherence time
 for a generic type of system-environment interaction
 and for a broad class of environments.
 This contrasts the situation of the master-equation approach with which only particular
 types of interaction and environment have been treated \cite{PZ99} while extension
 to more general situations is mathematically difficult.

 Let us assume that the environment is modelled by a quantum chaotic system \cite{note-Sred}.
 (Analogous strategy can be applied when the environment has a regular
 or mixed-type phase space structure.)
 Fidelity decay in such systems has been studied
 via semiclassical methods as well as with random
 matrix theory, both giving consistent results.
 Specifically, it turns out that for initial states chosen randomly,
 the fidelity has typically a Gaussian decay
below a perturbative border $\epsilon_p$ and an exponential decay
above this border \cite{fid-chao,CT02,wang}. If we model the environment $H_{\E}$ by a single
matrix of dimension $N$ taken from the so-called Gaussian orthogonal
ensemble (GOE) \cite{note},
 then the border $\epsilon_p $ can be explicitly estimated and is given by \cite{CT02}
 \be 2\pi \epsilon_p \overline{ V_{nd}^2} \sim \sigma_v \Delta , \label{per-bor} \ee
 where  $\overline{ V_{nd}^2}$ is the average of $|\la n|V|n'\ra |^2$
  with $n \ne n'$ and $V$ given by Eq.~(\ref{V-define}). Here
 $|n\ra $ denotes the eigenstates of
 $H^{eff}_{\E \alpha }, $ $\Delta $ is the mean level spacing of $H^{eff}_{\E \alpha }$,
 and $\sigma_{v}^2$ is the variance of $\la n|V|n\ra $.

 Below the perturbative border, $\epsilon < \epsilon_p$, the fidelity amplitude decays as \cite{CT02}
 \be |f_{\beta \alpha }(t)| \simeq e^{-\epsilon^2 \sigma^2_{v}t^2/2 \hbar^2}. \label{Gaussian} \ee
 {Then, the decoherence time $\tau_{d}$, characterizing the decay of off-diagonal
 matrix elements [see Eq.(\ref{off-rho})] is given by:}
 \be \label{tau-f1} \tau_d \simeq \sqrt{2} \hbar / (\epsilon \sigma_{v})  \propto \epsilon^{-1},
 \hspace{1cm} \epsilon < \epsilon_p . \ee
 Notice that this dependence of $\tau_d$ on $\epsilon$ coincides with the one derived
 in Ref.~\cite{PZ99}, even though in our case we do not assume an adiabatic  environment.
 For $\epsilon > \epsilon_p$, $|f_{\beta \alpha }(t)|$ has an exponential decay \cite{fid-chao}:
\bey |f_{\beta \alpha }(t)| \sim e^{-\Gamma t/ 2\hbar } \ \
\text{with} \  \Gamma =2\pi \epsilon^2 \overline{ V_{nd}^2}/ \Delta ,
 \label{ft-fgr}
 \\ \text{and} \hspace{1cm}
 \tau_d \simeq \hbar \Delta /[\pi \epsilon^2 \overline{ V_{nd}^2}] \propto \epsilon^{-2},
 \hspace{0.8cm} \epsilon > \epsilon_p. \label{dt-fgr} \eey
  {We stress that, like Eq.~(\ref{off-rho}), also Eqs.~(\ref{tau-f1}) and
  (\ref{dt-fgr}) are valid only if the time scale under consideration
  is much less than $\tau_E$.}

\begin{figure}[!t]
 \includegraphics[width=\columnwidth]{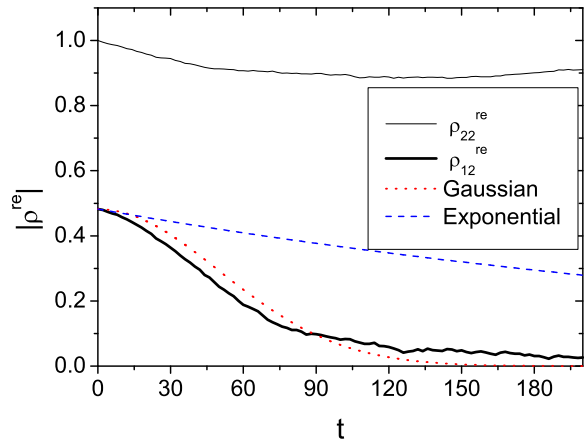}
   \vspace{-0.2cm}
      \caption{
          Same as in Fig.~\ref{fig-1} but  for $\epsilon =0.01$ which is still below but close to the perturbation
          border. It is seen that the numerically
          computed $|\rho^{re}_{12}|$ begin to deviate
               from the predicted Gaussian decay.
           Moreover the population decay (thin solid curve) becomes appreciable.
                 } \label{fig-2}
                    \end{figure}

 The next issue is if, and under what conditions, $\tau_E$ is sufficiently
 large so that significant decoherence may occur for $t\ll \tau_E$.
 If this is the case,  then
 energy eigenstates are much more robust than their superposition states.
 Let $|\mu_{\E}\ra $ be one eigenstate of $H_{\E}$
 and $ \la H^2_{I,nd} \ra $ be the mean square of the non-diagonal matrix elements
 $\la \alpha '|\la \mu_{\E}'| H_I |\mu_{\E}\ra |\alpha \ra $ ($\alpha\ne\alpha'$).
 One may estimate $\tau_E$ by using Fermi's golden rule,
 i.e.,
 \be \tau_E \simeq  1/R  \propto \epsilon^{-2}, \hspace{0.5cm} \text{where} \
 R = 2\pi \epsilon^2 \rho \la H^2_{I,nd} \ra / \hbar \label{te-fgr} \ee
 is the transition rate in Fermi's golden rule
 with the on-shell density-of-states $\rho$ approximated by
 the average density of all possible final states for the whole system \cite{Note2}.

 Therefore, below the perturbative border ($\epsilon<\epsilon_p$)
 the decoherence time $\tau_d$ and the relaxation time $\tau_E$ scale as
 $\epsilon^{-1}$ and $\epsilon^{-2}$, respectively. It follows that for small enough $\epsilon$,
   we have $\tau_d \ll \tau_E $  and hence
energy eigenstates are good preferred states
 regardless of the form of the coupling.
 On the other hand, above the perturbative border, both $\tau_d$
 in Eq.(\ref{dt-fgr}) and $\tau_E$ in Eq.(\ref{te-fgr}) scale with $\epsilon^{-2}$.
 In particular, in cases of very small $\epsilon_p$,
 the two time scales can be comparable \cite{T2T1note} even at weak perturbation and
 hence energy eigenstates may not be   preferred states .

We will now introduce a simple dimensionless model which will also
allow an explicit numerical evaluation of the different time scales.
Let $S$ be a qubit system with Hamiltonian
 $H_S = \sum_{\alpha} E_{\alpha}|\alpha \ra \la \alpha |$,
 $\alpha =1,2$ and $E_2-E_1=1$.
The environment is modeled by a $N$ dimensional matrix in the GOE
and the interaction $H_I$ is taken in the form of a random matrix as
well. Specifically, in a given arbitrary basis  $|k\ra $, the matrix
elements $\la k|H_{\E}|k'\ra $ and $\la \alpha k |H_I|\alpha 'k'\ra
$ are real random numbers distributed according to a Gaussian with
unit variance. Moreover we set the Planck constant $\hbar =1$.

  \begin{figure}[!t]
   \includegraphics[width=\columnwidth]{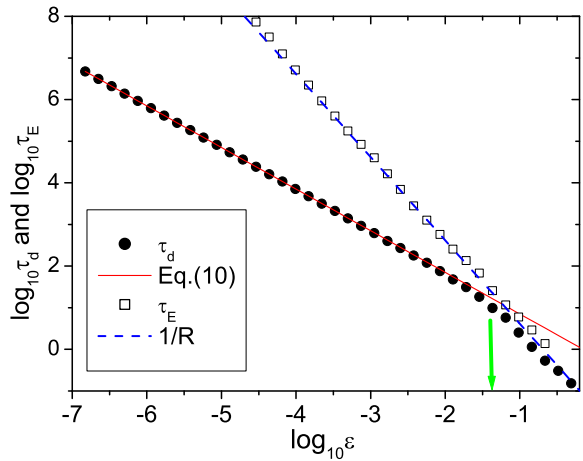}
     \caption{
      Dependence of times $\tau_d$ and $\tau_{E}$ on the
       coupling strength $\epsilon $.
     Full circles and empty squares represent the numerically computed $\tau_d$ and $\tau_E$,
     respectively.
The solid line is given by the theoretical expression
Eq.~(\ref{tau-f1}) and
    the dashed line is given by Eq.~(\ref{te-fgr}).
    Notice the good agreement between numerical data and  theoretical
    predictions.
The arrow indicates the perturbative border $\epsilon_p$ which
clearly separates
 the two different scaling behaviors of $\tau_d$.
         } \label{fig-te-tau}
              \end{figure}

We have numerically integrated the time-dependent Schr\"{o}dinger
equation for this model thus obtaining the matrix elements of the
reduced density matrix $\rho^{re}$. In Fig.~\ref{fig-1}, we show the
numerical results for parameters $N=200$ and $\epsilon =5\times
10^{-4}$. The perturbative border can be numerically computed from
Eq.~(\ref{per-bor}) and it is found to be $\epsilon_p \sim 0.04$. It
is clearly seen that for weak coupling the decay of $\rho^{re}_{12}$
is well predicted by the Gaussian decay in Eq.~(\ref{Gaussian}). By
contrast, the change in the diagonal matrix elements of the reduced
density is negligible, as shown by the upper thin curve in
Fig.~\ref{fig-1}.

Figure \ref{fig-2} is drawn for the same parameters of
Fig.~\ref{fig-1} but for a larger coupling strength, comparable to
the perturbative border. One notices  deviations from
 the Gaussian decay and also an appreciable population change
 (upper thin curve).
Interestingly, we found that deviations from the Gaussian decay
always goes with an appreciable population change. Indeed if, for
example, we deliberately weaken those off-diagonal coupling terms
(diagonal coupling terms not touched) that are responsible for the
population decay, then the Gaussian decay can be recovered while the
population decay becomes negligible.

Finally, if we further increase the coupling strength  above the
perturbative border, then the exponential decay of the off-diagonal
matrix elements [Eq.~(\ref{ft-fgr})] is indeed observed (not shown
here).

We have carefully studied the scaling behavior of $\tau_E$ and
$\tau_{d}$ as functions of the coupling strength $\epsilon $  and the
  results are shown in Fig.~\ref{fig-te-tau}. Numerically $\tau_{d}$ is defined
   as the time scale over which $|\rho^{re}_{12}|$ decays by a factor of $1/e$,
 and $\tau_E$ is defined as the reciprocal of the slope of $\rho^{re}_{22}(t)$
 in the initial interval of time in which Fermi's golden rule is valid.
 It is seen that numerical data nicely agree with analytical predictions.
 In particular, one can distinguish between the two scaling behaviors of $\tau_d$ ($\epsilon^{-1} $
 and $\epsilon^{-2}$) separated by the perturbative border $\epsilon_p$.
 Below this border, $\tau_{d}\ll \tau_E$,
 implying that energy eigenstates are much more stable than energy superposition states.

\vspace{0.5cm}
 \section{Discussions and Conclusions}

Entanglement-induced decoherence within a closed total system is
also referred to as ``intrinsic decoherence" \cite{gongpra}. The
above analysis directly indicates the existence of preferred states
of intrinsic decoherence.  As such, the dynamics of entanglement
generation between two subsystems depends strongly on the coherence
properties of the initial state. This is of interest to studies of
entanglement generation in classically chaotic systems \cite{FMT03}.
Further, our results might also shed light on the dynamics of
quantum thermalization processes within a closed system
 \cite{quantum-thermal}, where energy eigenstates also play a special role.

In summary, for weak but generic coupling between two quantum
subsystems, energy eigenstates are shown to play a special role in
the entanglement-induced decoherence process. The quality of the
energy eigenstates as preferred states is also analyzed
in terms of a simple dynamical model which allows for numerical analysis.

ACKNOWLEDGMENTS. The authors are very grateful to F.~Haake for
valuable discussions and suggestions. This work is supported in part
by an Academic Research Fund of NUS and the TYIA of DSTA, Singapore
under project agreement No.~POD0410553.
 W.W.~ is partially supported by Natural Science
Foundation of China Grant No.~10275011 and 10775123.
 J.G.~is supported by the
start-up funding and the NUS ``YIA" funding (WBS grant No.
R-144-000-195-123). C.G.~is funded by the MIUR-PRIN 2005 ``Quantum
computation with trapped particle arrays, neutral and charged''.


\begin{thebibliography}{99}

 \bibitem{Zurek81} W.~H.~Zurek, Phys.~Rev.~D {\bf 24}, 1516 (1981);
 W.~H.~Zurek, S.~Habib, and J.~P.~Paz, Phys.~Rev.~Lett.~{\bf 70}, 1187 (1993).

 \bibitem{BHS01} D.~Braun, F.~Haake, and W.T.~Strunz, Phys. Rev. Lett.~{\bf 86}, 2913 (2001);
 L.~Di\'{o}si and C.~Kiefer, Phys.~Rev.~Lett.~{\bf 85}, 3552 (2000);
 J.~Eisert, Phys.~Rev.~Lett.~{\bf 92}, 210401 (2004);
 H.~M.~Wiseman and J.~A.~Vaccaro, Phys.~Rev.~A {\bf 65}, 043606 (2002).

\bibitem{PZ99} J.~P.~Paz and W.~H.~Zurek, Phys.~Rev.~Lett.~{\bf 82}, 5181 (1999).


 \bibitem{fid-chao} R.A.~Jalabert and H.M.~Pastawski, Phys.~Rev.~Lett.~{\bf 86},
     2490 (2001); Ph.~Jacquod, P.G.~Silvestrov, and C.W.J.~Beenakker,
                  Phys.~Rev.~E {\bf 64}, 055203(R) (2001);
T.~Prosen and M.~\v{Z}nidari\v{c}, J.~Phys.~A {\bf 35}, 1455 (2002);
          F.M.~Cucchietti, {\it et al},
                  \pre{\bf 65}, 046209 (2002).
 \bibitem{CT02} N.~R.~Cerruti and S.~Tomsovic, J.~Phys.~A {\bf 36}, 3451 (2003);
  Phys.~Rev.~Lett. {\bf 88}, 054103 (2002).
\bibitem{wang} W.-G Wang and B.~Li, Phys.~Rev.~E {\bf 66}, 056208 (2002);
 W.-G.~Wang, G.~Casati, and B.~Li, {\it ibid}.~\textbf{69}, 025201(R)(2004);
  W.-G.~Wang, {\it et al}, {\it ibid}.~{\bf 71}, 037202 (2005);
   W.-G Wang and B.~Li, {\it ibid}.~{\bf 71}, 066203 (2005).
 \bibitem{GPSZ06} T.~Gorin, T.~Prosen, T.H.~Seligman, and M.~\v{Z}nidari\v{c},
 Phys.~Rep.~{\bf 435}, 33 (2006).
 \bibitem{CDPZ03} F.M.~Cucchietti, {\it et al}, Phys.~Rev.~Lett.~{\bf 91},
 210403 (2003); F.M.~Cucchietti, {\it et al}, Phys.~Rev.~A {\bf 72}, 052113 (2005).

\bibitem{GPSS04} T.~Gorin, T. Prosen, T.H. Seligman, and W.T. Strunz, \pra{\bf 70}, 042105 (2004);
 H.T. Quan {et al}., \prl{\bf 96}, 140604 (2006).

 \bibitem{note-Sred} M.~Srednicki, Phys.~Rev.~E {\bf 50}, 888 (1994).

 \bibitem{note} Decoherence effects with environments modeled by 
 random matrices
are recently considered in C. Pineda and T.H. Seligman, \pra{\bf 75}, 012106 (2007).

\bibitem{Note2} The same scaling with $\epsilon $ is derived in a more rigorous
treatment for $M$-level quantum systems interacting
with reservoirs in the thermodynamic limit
in M. Merkli, I.M. Sigal, and G.P. Berman, \prl{\bf 98}, 130401 (2007).
We also remark that the decoherence time given in that paper also scales as
$\epsilon ^{-2}$, where the initial state of the environment is
assumed to be an equilibrium state
 and the off-diagonal elements of reduced density matrix are studied with respect to their
 equilibrium values.
 \bibitem{T2T1note} {Similar to the well-known $T_{2}$-$T_{1}$
 relation, the relaxation time scale $\tau_{E}$} sets an upper bound to $\tau_{d}$.
\bibitem{gongpra}J.~Gong and P.~Brumer, Phys.~Rev.~A {\bf 68}, 022101 (2003).
 \bibitem{FMT03}
 H.~Fujisaki, T.~Miyadera, and A.~Tanaka, Phys.~Rev.~E {\bf 67}, 066201 (2003);
 R.~D.-Dobrza\'{n}ski and M.~Ku\'{s}, Phys.~Rev.~E {\bf 70}, 066216 (2004);
K. Furuya, M.C. Nemes, and G.Q. Pellegrino, \prl{\bf 80}, 5524
(1998).
\bibitem{quantum-thermal}S. Popescu, A.J. Short, and A.
Winter, Nature Physics {\bf 2}, 754 (2006); S. Lloyd, Nature Physics
{\bf 2}, 727 (2006); S. Goldstein, {\it et al}, \prl{\bf 96}, 050403 (2006).



 \end{thebibliography}
 \end{document}